\documentstyle[12pt,epsf,here]{article}

\def \be {\begin{equation}}
\def \e {\end{equation}}
\def \bea {\begin{eqnarray}}
\def \ea {\end{eqnarray}}
\def \ep {\epsilon}

\def \no {\nonumber}

\newcommand{\To}[2]{\stackrel{#1}{\hbox to #2 pt{\rightarrowfill}}}
\def \an {\widehat}

\def \vector#1{\stackrel{\hspace{-0.45em}\longrightarrow}{#1}}

\def\np#1#2#3{{\it Nucl.~Phys.\/}~{\bf B#1} (19#2) #3}
\def\pl#1#2#3{{\it Phys.~Lett.\/}~{\bf B#1} (19#2) #3}

\def\prd#1#2#3{{\it Phys.~Rev.\/}~{\bf D#1} (19#2) #3}
\def\prl#1#2#3{{\it Phys.~Rev.~Lett.\/}~{\bf #1} (19#2) #3}

\def\epj#1#2#3{{\it Eur.~Phys.~J.\/}~{\bf C#1} (19#2) #3}

\begin{document}  
\vspace*{-2cm}  
\renewcommand{\thefootnote}{\fnsymbol{footnote}}  
\begin{flushright}  
hep-ph/9905341\\
DTP/99/42\\  
May 1999
\end{flushright}  
\vskip 65pt  
\begin{center}  
{\Large \bf Radiation Zeros in \boldmath $W^+W^-\gamma$ Production
 \\[3mm]
at High-Energy Colliders  } \\ 
\vspace{1.2cm} 
{\bf  

W.~James~Stirling${}^{1,2}$\footnote{W.J.Stirling@durham.ac.uk}  and
Anja Werthenbach${}^1$\footnote{Anja.Werthenbach@durham.ac.uk} }\\  
\vspace{10pt}  
{\sf 1) Department of Physics, University of Durham,  
Durham DH1 3LE, U.K.\\  
  
2) Department of Mathematical Sciences, University of Durham,  
Durham DH1 3LE, U.K.}  
  
\vspace{70pt}  
\begin{abstract}
The vanishing of the cross section for particular points in phase space
 -- radiation zeros -- is examined for the process 
 $ q \bar q \to W^+W^- \gamma$
at high energy.  Unlike the process $q \bar q\, {'}\to W^\pm \gamma$, actual
zeros only occur in the soft-photon limit. However, for photon energies 
that are not too large, the cross section does  exhibit deep dips in
 regions of phase space corresponding to the position of the actual zeros. We show  that in these regions the sensitivity
to possible anomalous quartic couplings is very large.
\end{abstract}
\end{center}  
\vskip12pt

\setcounter{footnote}{0}  
\renewcommand{\thefootnote}{\arabic{footnote}}  
  
\vfill  
\clearpage  
\setcounter{page}{1}  
\pagestyle{plain} 

\section{Introduction}

In  certain   high--energy   scattering  processes  involving  charged
particles  and the  emission of one or more  photons,  the  scattering
amplitude  vanishes for particular  configurations of the final--state
particles.  Such  configurations are known as {\it radiation zeros} or
{\it null zones}.
The study of these radiation zeros (RAZ) dates back to  
the late 1970s  \cite{first}, where they were identified in the process 
 $ q \bar{q}{'} \to W \gamma$ as points in phase-space for which the total cross
section vanishes.\\ 

Today \cite{brown} it is understood that the zeros are due to a cancellation which 
can be regarded as a destructive
interference of radiation patterns induced by  the charge of the participating 
particles. The fact that gauge symmetry is a vital ingredient for the cancellation
to occur means that  radiation zeros can be used to probe physics beyond the standard model.
For example, `anomalous' electroweak gauge boson couplings destroy
the delicate cancellations necessary for a zero to occur. \\

In recent years there have been many  studies  exploring  the  phenomenological
aspects  of  radiation   zeros, see for example Ref.~\cite{brown} and references therein.
Experimental  evidence for the zeros predicted in \cite{first}
 has also been found at the Fermilab Tevatron $p \bar
p$  collider   \cite{cdf}. \\

As already mentioned, the classic process for radiation zeros
in high-energy hadron-hadron collisions is $ q \bar{q}{'} \to W \gamma$, where
the zero occurs in a `visible' region of phase space, i.e. away from the phase-space
boundaries. It is natural to extend the analysis
to more complicated processes involving multiple gauge boson production. At the upgraded Tevatron $p \bar p $ and LHC
$p p$ colliders, the rates for such events can be quite large.  \\

In this paper we study in detail the  $ q \bar{q} \to W W\gamma$ process, and identify
the circumstances under which radiation zeros occur. Unlike the $  W\gamma$ process,
it is not possible to write down a simple analytic expression for the matrix element
squared. However, making use of the soft-photon approximation does
allow the zeros to be identified analytically, and a numerical calculation of the full
matrix element confirms that although the actual zeros disappear for non-zero photon energies, deep dips do persist for all relevant photon energies.  The dips result from
delicate cancellations between the various standard model photon emission diagrams,
and are `filled in' by contributions from non-standard gauge boson couplings. We illustrate 
this explicitly using anomalous quartic couplings.\\

The paper is organised as follows. In the following section we review the `classic'
radiation-zero process, $ q \bar q {'}\to W \gamma$. In Section~3 we extend the analysis
to $W^+W^-\gamma$ production, first using analytic methods in the soft-photon limit.
We carefully distinguish between photons emitted in the $W^+W^-$ production process
and those emitted in the $W\to f \bar f$ decay process. We  then extend 
the analysis to non-soft photons using a numerical calculation of the exact matrix 
element. In Section~4 we show how anomalous quartic couplings `fill in' the dips
caused by the radiation zeros. Finally, Section~5 presents our summary and conclusions.
\\

\section{Radiation Zeros in \boldmath $W\gamma$ Production}

\begin{figure}[H]
\vskip -2.8cm
\hspace{1.3cm}\centerline{\epsfysize=6cm\epsffile{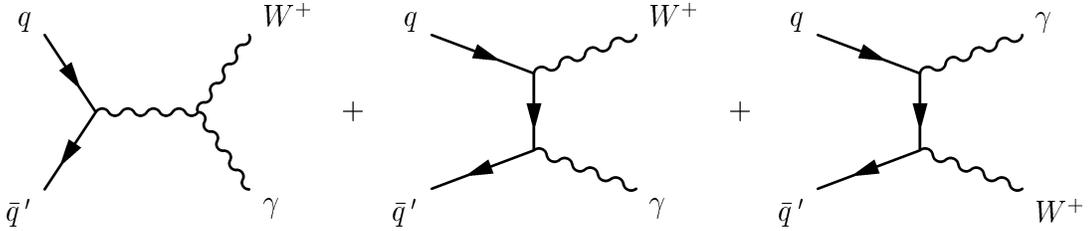}}
\caption{\label{single} {Diagrams contributing to the process $ q \bar{q}{\, '} \to W^+ \gamma$.}}
\end{figure}

The classic scattering process which exhibits a radiation zero 
is $ q \bar{q} \, {'} \to W^+ \gamma$. The amplitude for this can 
be calculated analytically  --- there are three
Feynman diagrams, shown in Fig.~\ref{single}. \\

With momenta labelled as
\bea
q(p_1) + \bar{q} \, {'}(p_2) & \to &  W^+(k^+) + \gamma(k) \nonumber \\
W^+(k^+) & \to &  \nu_l(r_3) + l^+(r_4)  \; ,
\ea
the matrix element is
\bea
{\cal M} & = &    \, \frac{-i e g}{\sqrt{2\, p_2 \cdot k} } 
\; {\cal C} \; \left\{ t(r_3,p_2)\,s(k,p_1) \, [ s(p_1,r_4)\, t(p_1,p_2) \,+\, s(k,r_4)\, t(p_2,k)]
\right\} \no \\
{\cal C}& = & 
\frac{1}{p^2-M_{W}^2 } \, + \, \frac{Q_q}{(p_1-k)^2} \; ,
\label{eqamp}
\ea
where $p=p_1+p_2 = k^+ + k$.
Here we have used the spinor technique of Ref.~\cite{kleiss}, with photon polarisation vector\footnote{The expression in Eq.~(\ref{eqamp}) actually
corresponds to a positive helicity photon. For a negative helicity photon, a similar expression is obtained.
Both amplitudes exhibit the same radiation zero.} $\epsilon_{+ \mu}^{\phantom{+}\ast}(k)=(1/\sqrt{4 \, p_2 \cdot k})\,\, \bar{u}_+(k) \gamma_{\mu} u_+(p_2)$. The spinor products are defined by
\be
s(p_i,p_j)  =  \bar{u}_+(p_i)u_-(p_j) , \qquad t(p_i,p_j) =  \bar{u}_-(p_i) u_+(p_j)\; ,
\e
and all fermion masses are set to zero. \\

The cross section $  \sigma \sim |{\cal M}|^2$ therefore vanishes when ${\cal C} = 0$, i.e.
\be
\label{solve}
\frac{1}{p^2-M_{W}^2 } \, = \, \frac{Q_q}{2 \, p_1 \cdot k} \; .
\e
We next introduce the momentum four-vectors 
\bea
p_1^{\phantom{1}\mu} \, &=& \, ( E, 0, 0, E) \no \\
p_2^{\phantom{2}\mu} \, &=& \, ( E, 0, 0, -E) \no \\
k_+^{\phantom{+}\mu} \, &=& \, \left( \frac{4\, E^2+M_{{W}}^2}{4\, E}, \frac{4\, E^2-M_{W}^2}{4\, E}\,
 \sin \, \Theta \, , \, 0 \, , \, \frac{4\, E^2-M_{W}^2}{4\, E} \cos \, \Theta \right) \no \\
k^{\mu} \phantom{+} \, &=& \, \left( \frac{4\, E^2-M_{W}^2}{4\, E}, - \frac{4\, E^2-M_{W}^2}{4\, E}\,
 \sin \, \Theta \, , \, 0 \, , \, -\frac{4\, E^2-M_{W}^2}{4\, E} \cos \, \Theta \right)\; ,\no \\
\ea
where $ \Theta $ is the angle between the incoming quark and the $W^+$, $\theta_{\gamma} = \Theta +\pi $, 
and $E$ is the beam energy of the scattering particles. Substituting into Eq.~(\ref{solve}) gives the condition
for a radiation zero \cite{first}:
\bea
 \cos \, \Theta \, &=& \, -1 \, + \, 2 \, Q_q \; .
\ea
In other words, the cross section vanishes when the photon is produced at an angle\footnote{A similar condition holds for the process $q\bar{q} \, {'} \to W^-\gamma$: $\cos \theta _{\gamma} \,^{\scriptscriptstyle {RAZ}} = 1+2Q_q =\frac{1}{3}, $ for $q=d$.}
\be
\label{raz}
\cos \theta _{\gamma} \,^{\scriptscriptstyle {RAZ}} = 1-2Q_q = 
-\frac{1}{3} \qquad \mbox{for $q=u$} \; .\\  
\e

The angle $\theta_\gamma$ for which the cross section vanishes is independent of the photon
energy, in particular it is unchanged in the soft-photon limit, $k^\mu \to 0$, which is
realised as the beam energy decreases to its threshold value, $2E \to M_{ W}$.
In this limit we can use the {\it eikonal approximation} to locate the position of the zero.
Since for more complicated processes we may only be able to obtain an analytic
expression in this approximation, it is worth repeating the above calculation in
the soft-photon limit to check that we do indeed obtain the same result.\\

We start from the matrix element for the process $ q \bar{q}{'} \to W^+$: 
\be
i {\cal M}_0 = \bar{u}_-(p_2) \, \left( i \frac{g}{\sqrt{2}} \right)  \gamma^{\nu} \, u_-(p_1)\, 
\ep_{\nu}^* (k_+) \; .
\e
In the soft-photon limit one can neglect the momentum $k$ in the numerators of the 
internal fermion propagators,  
in the $WW\gamma$  vertex, and in the overall energy-momentum conservation constraint
(i.e. $ p=k_+ =  p_1+p_2$), which leads to 
\bea
\label{m}
{\cal M} = (-e) \; {\cal M}_0 \; \ep_{\mu}^* (k) \;j^{\mu}
\ea
where the eikonal factor $j^{\mu}$ is given by
\be
\label{eikonal}
 j^{\mu} \, = \, \, Q_q \, \frac{ p_1 ^{\phantom{1}\mu}}{p_1 \cdot k} \, + \, (1-Q_q)\,  
\frac{ p_2 ^{\phantom{2}\mu}}{p_2 \cdot k} 
\, - \, \frac{ k_+ ^{\phantom{+}\mu}}{k_+ \cdot k}  \; .
\e
The three terms in $j^\mu$ come from the $u-$, $t-$ and $s-$channel diagrams respectively or,
equivalently, a soft photon radiated off the incoming quark, incoming antiquark, 
and outgoing $W^+$. Note that gauge invariance implies $k_{\mu}\cdot j^\mu\, =\, 0$.\\

Radiation zeros are now obtained for  $\ep^*(k) \cdot j \, = \, 0$. Choosing 
\bea
k^{\mu} &=& E_{\gamma}\, (1, \sin \, \theta_{\gamma} ,0, \cos \, \theta_{\gamma}) \no \\
\ep^{\mu }_1 &=&  (0,0,1,0) \no \\
\ep^{\mu }_2 &=&  (0, -\cos \, \theta_{\gamma}, 0, \sin \, \theta_{\gamma})
\label{polvecs}
\ea
gives
\be
\ep^*(k) \cdot j \, = \, - Q_q \, 
\frac{\sin \, \theta_{\gamma}}{1- \cos \, \theta_{\gamma}} \, + \, (1-Q_q) \, 
 \frac{\sin \, \theta_{\gamma}}{1+ \cos \, \theta_{\gamma}} \, = \, 0 \; ,
\e
or equivalently
\be
\cos \theta _{\gamma} \,^{\scriptscriptstyle {RAZ}} = 1-2Q_q \; .
\e
This is exactly the same condition as Eq.~(\ref{raz}), as expected.
\\


\section{Radiation Zeros in  \boldmath $W^+W^-\gamma$ Production}

\begin{figure}[H]
\vskip -1.8cm
\hspace{1.1cm}\centerline{ \epsfysize=20cm\epsffile{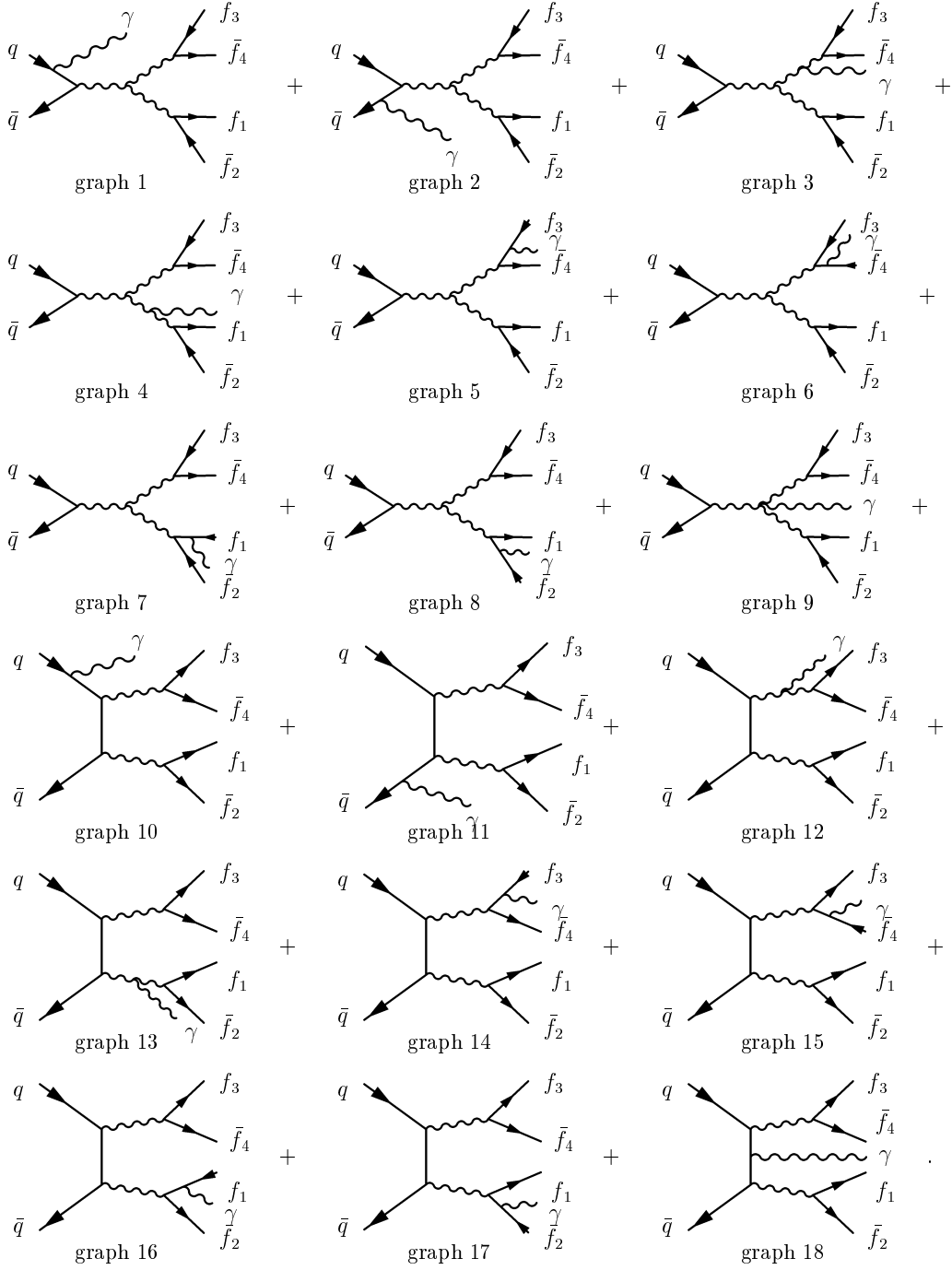}}
\caption{\label{all} Feynman diagrams for the process 
$ q \bar{q} \to W^+W^-\gamma \to f_1 \bar{f_2} f_3 \bar{f_4} 
\gamma$.}
\end{figure}

In this section we extend the analysis to investigate radiation
zeros in the process $ q \bar{q} \to W^+W^-\gamma \to f_1 \bar{f_2} f_3 \bar{f_4} \gamma$. 
The contributing Feynman diagrams are shown in Fig.~\ref{all}.
Note that both $\gamma$ and $Z$ exchange are included in the $s-$channel diagrams. \\

We first calculate the matrix element in the soft-photon approximation.
Once again the matrix element can be factorized:
\be
\label{M}
{\cal M} = (-e) \; {\cal M}_0 \; \ep_{\mu}^* (k) \;j^{\mu}
\e
where  ${\cal M}_0$ is  the ($q \bar{q}\to W^+W^-$) 
matrix element without photon radiation, and the eikonal current is
\bea
\label{Eikonal}
\quad j^{\mu} &=& \quad \left(  Q_3 \, \frac{r_3^{\phantom{3}\mu}}{r_3 \cdot k} \, + \, (1-Q_3) \, 
\frac{r_4^{\phantom{4}\mu}}{r_4 \cdot k} \, - \, \frac{k_+^{\phantom{+}\mu}}{k_+ \cdot k}\right)\, 
\frac{k_+^2-\overline{M}^2}
{(k_++k)^2-\overline{M}^2} \qquad \quad \qquad \no \\
 && - \left(  Q_1 \, \frac{r_1^{\phantom{1}\mu}}{r_1 \cdot k} \, + \, (1-Q_1) \, 
\frac{r_2^{\phantom{2}\mu}}{r_2 \cdot k} \, - \, \frac{k_-^{\phantom{-}\mu}}{k_- \cdot k} \right)\, 
\frac{k_-^2-\overline{M}^2}
{(k_-+k)^2-\overline{M}^2} \no \\
 && + \left(- Q_q \, \frac{p_1^{\phantom{1}\mu}}{p_1 \cdot k} \, - \, Q_{\bar{q}} \, \frac{p_2^{\phantom{2}\mu}}{p_2 
\cdot k} \, + \, \frac{k_+^{\phantom{+}\mu}}{k_+ \cdot k} \, - \, \frac{k_-^{\phantom{-}\mu}}{k_- \cdot k} \right) \no 
\\
&& \no \\ 
 &=& \qquad D^{+\mu} \, -\, D^{-\mu} \, +\, P^{\mu}  
 \ea  
with $Q_i= |Q_i|> 0$, $Q_{\bar{q}}=-Q_q$, $r_i$ the momenta of the final-state $f_i$ fermions 
and $ \overline{M } = M_{ W} - i \, \Gamma_W /2 $. This result is appropriate
for both right-handed and left-handed quark scattering, although 
 ${\cal M}_0$ is of course different in the two cases.\\
 
\begin{figure}[H]
\vspace{-3cm}
\hspace{0.4cm}\centerline{\epsfysize=6cm\epsffile{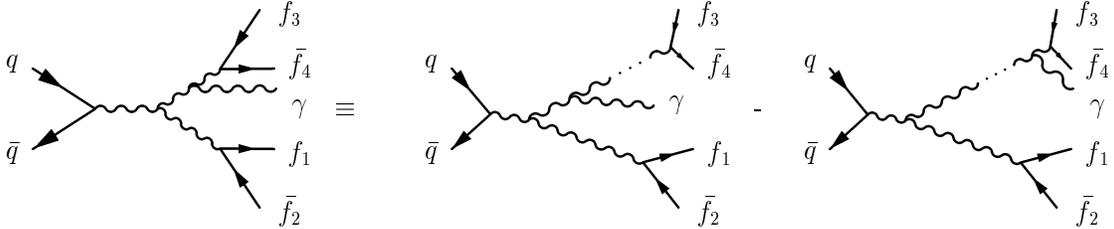}}
\caption{ \label{parfrac} Partial fractioning of photon emission off a final-state $W$ boson.}
\end{figure}

In deriving Eq.~(\ref{Eikonal})  we have made use of the partial fraction 
\bea
\label{partial} 
\frac{1}{k_{\pm}^2-\overline{M}^2}\, \frac{1}{(k_{\pm}+k)^2-\overline{M}^2} = \frac{1}{2\, k_{\pm} \cdot k} \, \left( \, 
\frac{1}{k_{\pm}^2-\overline{M}^2}\, - \, \frac{1}{(k_{\pm}+k)^2-\overline{M}^2} \right) \qquad
\ea
to split the contributions involving  photon emission from the final-state
$W$ bosons into two pieces corresponding to photon emission {\it before}
and {\it after} the boson goes on mass shell \cite{tt}.
This is illustrated in Fig.~\ref{parfrac}. \\

%
%
To obtain the cross section one has to integrate over the virtual momenta $ k_{\pm}$:
\be 
\sigma \sim \, \int dk_+^2\, dk_-^2 \, \sum \, | {\cal M}|^2 \, \simeq \, \left( \frac{\pi}{M_{{ W}}\, \Gamma_W}
 \right)^2 \, \sum \, | {\widetilde{\cal M}}_0|^2 \, e^2 \, {\cal F}
\e
with 
\bea
{\cal M}_0 &=& {\widetilde{\cal M}}_0 \, \frac{1}{k_+^2-\overline{M}^2} \, \frac{1}{k_-^2-\overline{M}^2} \no \\
{\cal F} &\equiv& \left( \frac{M_{{ W}}\, \Gamma_W}{\pi} \right)^2 \, \int dk_+^2\, dk_-^2 \, ( - j \cdot j^*) \, 
\frac{1}{|k_+^2-\overline{M}^2|^2} \, \frac{1}{|k_-^2-\overline{M}^2|^2} \; .
\ea
Performing the $k_\pm^2$ integrals by completing the contours in an appropriate half plane and using 
Cauchy's theorem eventually leads to 
\bea
{\cal F} = |P|^2\, +\, |D^+|^2 \, + \, |D^-|^2 \, - \, 2 {\rm Re} \, [ D^+\, D^{-*}]\, +\, 2{\rm Re}\,
[P(D^{+*}- D^{-*})] \; ,
\ea
with
\bea
\label{big}
|P|^2\, &=& Q_q^2 \, \an{p_1 p_1} + Q_{\bar{q}}^2 \, \an{p_2p_2} + \,\an{k_+k_+} + \,\an{k_-k_-} + 2\, Q_q Q_{\bar{q}}\, 
\an{p_1p_2} \no \\
&& -2\, Q_q \,\an{p_1k_+} + 2\, Q_{\bar{q}} \,\an{p_2k_-} +2\, Q_q \,\an{p_1k_-} -2 \,Q_{\bar{q}} \,\an{p_2k_+} -2\, 
\,\an{k_+k_-} \no \\
&& \no \\
|D^+|^2 \, &=& Q_3^2 \, \an{r_3r_3} + (1-Q_3)^2 \, \an{r_4r_4} + \, \an{k_+k_+} + 2 \, Q_3 (1-Q_3) \, \an{r_3r_4} \no \\
&& -2\, Q_3\, \an{r_3k_+} -2\, (1-Q_3) \, \an{r_4k_+} \no \\
&& \no \\
|D^-|^2 \, &=&Q_1^2 \, \an{r_1r_1} + (1-Q_1)^2 \, \an{r_2r_2} + \, \an{k_-k_-} + 2 \, Q_1 (1-Q_1) \, \an{r_1r_2} \no \\
&& -2\, Q_1\, \an{r_1k_-} -2\, (1-Q_1) \, \an{r_2k_-} \no \\
&& \no \\
2 {\rm Re} \, [ D^+\, D^{-*}]\, &=& -2\, \chi_{+-} \Big( Q_1 Q_3 \, \an{r_1r_3} + Q_1 (1-Q_3)\, \an{r_1r_4} - Q_1 \, 
\an{r_1k_+} \Big. \no \\
&&   + (1-Q_1)Q_3 \, \an{r_2r_3}  +(1-Q_1)(1-Q_3) \, \an{r_2r_4} \no \\
&&  \Big. -(1-Q_1) \, \an{r_2k_+} - Q_3 \, \an{r_3k_-} -(1-Q_3)\, \an{r_4k_-} + \an{k_+k_-}
 \Big) \no \\
 && \no \\
 2{\rm Re}\,[P(D^{+*}- D^{-*})] &=&  2\,  \chi_+ \Big( -Q_q Q_3 \, \an{r_3p_1} - Q_{\bar{q}} Q_3 \, \an{r_3p_2} + 
 \an{k_+k_-} +Q_3 \, \an{r_3k_+} -Q_3\, \an{r_3k_-}\Big. \no \\
 &&  -(1-Q_3)Q_q\, \an{r_4p_1} - (1-Q_3) Q_{\bar{q}} \, \an{r_4p_2}- \an{k_+k_+} \no \\
 &&  \Big. + (1-Q_3) \, \an{r_4k_+} - (1-Q_3) \, \an{r_4k_-} + Q_q \, \an{k_+p_1} + Q_{\bar{q}} \, \an{k_+p_2} \Big) \no \\
 && \no \\
 && -2 \chi_- \Big( - Q_1 Q_q\, \an{r_1p_1} - Q_1Q_{\bar{q}} \, \an{r_1p_2} + Q_1\, \an{r_1k_+} + \an{k_-k_-}
 - Q_1\, \an{r_1k_-} \Big. \no \\
 &&  - (1-Q_1)Q_q \, \an{r_2p_1} - (1-Q_1) Q_{\bar{q}} \, \an{r_2p_2} + (1-Q_1) \, \an{r_2k_+} \no \\
  && \Big. - (1-Q_1) \, \an{r_2k_-} + Q_q \, \an{k_-p_1} + Q_{\bar{q}} \, \an{k_-p_2} - 
  \an{k_-k_+} \Big) \; .\no \\
\ea
The `antennae'  appearing in this expression are defined by  
\be
\an{p_1p_2} = \frac{p_1 \cdot p_2}{ p_1 \cdot k \,\,\, p_2 \cdot k}
\e
and the profile functions \cite{tt} by
\bea
\chi_{+-} &=& \frac{[(k \cdot k_+)\,( k \cdot k_-) + ( \Gamma_W M_{{ W}})^2 ] \, (\Gamma_W M_{{ W}})^2}{ [(k \cdot k_+)^2 
+ (\Gamma_W M_{{ W}})^2]\, 
[(k \cdot k_-)^2 + ( \Gamma_W M_{{ W}})^2]} \no \\
 && \no \\
\chi_+ &=& \frac{(\Gamma_W M_{{ W}})^2}{(k \cdot k_+)^2 + ( \Gamma_W M_{{ W}})^2} \no \\
 && \no \\
 \chi_- &=& \frac{(\Gamma_W M_{{ W}})^2}{(k \cdot k_-)^2 + ( \Gamma_W M_{{ W}})^2} \quad . \no \\
\ea
This result agrees with that given in Ref.~\cite{orr}, where the distribution of soft radiation
accompanying $W^+W^-$ production in $e^+e^-$ annihilation was studied.\\

The profile functions have two important limits that have to be distinguished carefully.

\bigskip

\noindent{\bf (a) } {\boldmath $ E_{\gamma} \ll \Gamma_W \ll 1 $}
\medskip

\begin{figure}[H]
\vskip -2cm
\centerline{ \epsfysize=12cm\epsffile{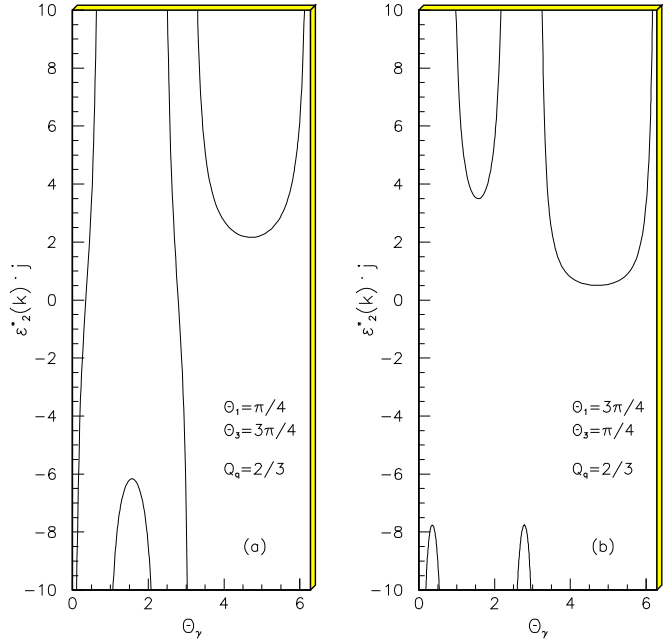}}
\vskip -1.8cm
\caption{\label{radpattern} {The radiation pattern of Eq.~(\ref{cot}). Two (a) or no (b) radiation zeros are visible.}}
\end{figure}

The photon is far softer than the $W$ is off mass shell, which leads to 
$\chi_{+-} = \chi_- = \chi_+ = 1$. The timescale for photon emission is much longer
than the $W$ lifetime, and so the photon `sees' only the external fermions.
The whole
current contributes and rather than solving ${\cal F} =0 $ to find radiation zeros 
we can determine the  values of $k^\mu$  for  which
\be
\label{ej}
\ep^*(k) \cdot j =0 \; .
\e
To simplify the calculation slightly we  consider only {\it leptonic} 
decays of the $W$s\footnote{The hadronic $W$ decay case, $W\to q \bar{q}'$
simply introduces a few extra terms, but the results are qualitatively unchanged.}.
  The eikonal current then reduces to\footnote{Note that in the soft limit the ratio of propagators in Eq.~(\ref{Eikonal}) is $\frac{k_{\pm}^2-\overline{M}^2}{(k_{\pm}+k)^2-\overline{M}^2} \to 1$.}
\be
j^{\mu} \, = \, \frac{r_3^{\mu}}{r_3 \cdot k} - \, \frac{r_1^{\mu}}{r_1 \cdot k} \, - Q_q \, \left(
 \frac{p_1^{\mu}}{p_1 \cdot k} - \frac{p_2^{\mu}}{p_2 \cdot k} \right)  \; .
\e
Here, $r_3$ is the four-momentum of the outgoing lepton with charge $+1$ and
$r_1$ is the four-momentum of the outgoing lepton with charge $-1$.
It turns out that the only solutions of $\ep^*(k) \cdot j =0$ occur when the scattering
is {\it planar}, i.e. all incoming and outgoing three-momenta lie in the same 
plane.\footnote{The planarity condition gives rise to the so-called Type II zeros discovered
recently \cite{james}.} If, as in Eq.~(\ref{polvecs}), we take one polarisation vector $\ep^*_1$ 
perpendicular
to this plane, and the other $\ep^*_2$ in the plane and orthogonal to the photon three-momentum,
then   $\ep^*_1(k) \cdot j =0$ is trivially satisfied 
and   $\ep^*_2(k) \cdot j =0$  leads to an implicit equation
for the   photon production angle $\theta_\gamma$  which corresponds to a radiation zero:
\be
\label{cot}
\cot \, \frac{\theta_{1\gamma}}{2} - \cot \, 
\frac{\theta_{3\gamma}}{2} - Q_q \left( \cot \, \frac{\theta_{\gamma}}{2} 
+ \tan \, \frac{\theta_{\gamma}}{2} \right) = 0  \; ,
\e
with $ \theta_{1\gamma}= \theta_1 - \theta_{\gamma}$ and $ \theta_{3\gamma}= 
\theta_3 - \theta_{\gamma}$ and where
the lepton four-momentum vectors are
\bea
r_1^{\mu} &=& E_1 (1, \sin \theta_1, 0, \cos \theta_1) \no \\
r_3^{\mu} &=& E_3(1, \sin \theta_3, 0, \cos \theta_3) \; .
\ea
Depending on the values of $\theta_1 $ and $\theta_3$, Eq.~(\ref{cot}) has either 
two solutions 
( $ \theta_1 > \theta_3 > \pi$ or $\theta_1 < \theta_3 < \pi$) or 
no solutions\footnote{One solution if either $\theta_1=\pi$ or $\theta_3=\pi$.}. This is illustrated in Figs.~\ref{radpattern}(a) and \ref{radpattern}(b) respectively.
The radiation pattern given by Eq.~(\ref{cot}) is plotted as a function of $\theta_\gamma$
for `typical' values of the lepton production angles, chosen such that the zeros 
(in the former case) are clearly visible.

\vspace{0.7cm}

The generalisation to the case of arbitrary $W$ decays is straightforward. Thus for 
\bea
W^- &\to& f_1(r_1, Q_1) + \bar f_2(r_2,Q_2 =1-Q_1) \no \\
W^+ &\to& f_3(r_3,Q_3) + \bar f_4(r_4,Q_4=1-Q_3) \; ,
\ea
Eq.~(\ref{cot}) becomes
\bea
\label{cotbis}
&& \hspace*{-2cm}-2Q_q \, \frac{1}{\sin \theta_{\gamma}} + Q_1 \frac{\sin \theta_{1\gamma}}{1- \cos \theta_{1\gamma}} - (1-Q_1) \frac{\sin \theta_{2\gamma}}{1+\cos \theta_{2\gamma}} \no \\
&& \hspace{1.5cm} -  Q_3 \frac{\sin \theta_{3\gamma}}{1-\cos \theta_{3\gamma}} + (1-Q_3) \frac{\sin \theta_{4\gamma}}{1+\cos \theta_{4\gamma}} \, = \,0   .
\ea
There are now either 4, 2 or 0 radiation zeros, depending on the relative orientation
in the plane of the initial- and final-state particles. \\

\noindent{\bf (b) }{\boldmath $ \Gamma_W \ll E_{\gamma} \ll 1 $}\\

When the photon is far harder in energy  than the $W$ is off mass shell (but still soft
compared to the $W$ masses and energies), 
the timescale for photon emission is much shorter than the $W$ lifetime. 
As far as the photon is concerned,
the overall process separates into `$W$ production' and `$W$ decay' pieces, with
no interference between them. Formally, in this limit the profile functions are
$\chi_{+-} = \chi_- = \chi_+ = 0$. Therefore all the interference terms in Eq.~(\ref{big})
 vanish, and to find zeros one has to solve
\be
{\cal F} = |P|^2 \, + \, |D^+|^2 \, + \, |D^-|^2 = 0  \; .
\label{factorise}
\e
Since each of these quantities is positive definite they have to vanish separately: 
\bea
\label{sep}
|P|^2 &=& 0 \:\:\: {\rm RAZ \:\: of } \:\:q \bar{q}\, \to\, WW\gamma \no \\
|D^+|^2 &=& 0 \:\:\:{\rm RAZ \:\: of } \:\: W^+ \, \to \, f_3 \bar{f_4} \gamma \no \\
|D^-|^2 &=& 0 \:\:\:{\rm RAZ \:\: of } \:\: W^- \, \to \, f_1 \bar{f_2} \gamma \quad .
\ea
Fortunately, the zeros of each are well-separated in phase space in regions that
can be isolated experimentally.  Thus in practice an energetic photon can be classified
as a `production' or a `decay' photon depending on whether it reconstructs to an invariant
mass $M_W$ when combined with the $W$ fermion decay products. 
Provided $E_{\gamma} \gg \Gamma_W$ this 
classification is in principle  unambiguous.
The radiation zeros for  $W^{\pm} \to f \bar f\,' \gamma$ decay have been known for some time,
and in fact are directly analogous to those for $q \bar q\,' \to W^{\pm} \gamma$ discussed 
in the previous section. \\

We therefore restrict our attention to the zeros of $q \bar q \to W^+W^-\gamma$,
given by $|P|^2 = 0 $, where the $W$s
are now considered {\it on-shell stable particles}. It is straightforward to 
derive the expression for the current in this case (cf. Eq.~(\ref{Eikonal})): 
\be
j^{\mu} \, = \,- \frac{k_-^{\mu}}{k_- \cdot k}
 + \, \frac{k_+^{\mu}}{k_+ \cdot k} \, - Q_q \, \left(
 \frac{p_1^{\mu}}{p_1 \cdot k} - \frac{p_2^{\mu}}{p_2 \cdot k} \right)  \; .
\e
Then solving $\ep^*(k)\cdot j = 0$ leads to
\bea
\label{tan}
\tan \theta_{\gamma} &=& \frac{ -\beta \, 
\sin\Theta - 2\, Q_q \beta^2 \cos \Theta \sin \Theta
\pm \sqrt{\beta^2 \, \sin ^2 \Theta + 4\, Q_q(\beta^2-1)\, 
( Q_q+\beta\, \cos \Theta)} }{2\,( -\beta \, 
\cos \Theta - Q_q + Q_q\, \beta^2 \, \sin ^2 \Theta)} \no \\
\ea
where $\beta=\left( 1-  M_{{W}}^2 / E^2 \right)^\frac{1}{2} < 1 $ is the velocity of the $W$,
$\Theta$ is the angle between the $W^-$ and the incoming quark, and $E$ is the beam energy. 
Note that again these results correspond to all incoming and outgoing particles lying
in the same plane. 
One interesting feature of this result is that there is now a 
certain minimum beam energy, for a given $\Theta$ and $Q_q$,
 which is required to set up the environment for radiation
zeros (the square root in Eq.~(\ref{tan}) has to be positive).  
For example, for $\Theta = \pi/2$ the critical energy is $E_{\mathrm crit.}= M_W(1+4Q_q^2)^{\frac{1}{2}}$. For energies $E > E_{\mathrm crit.}$ four radiation zeros are present (due to the $\pm$ and the periodicity of $\tan$). For $E = E_{\mathrm crit.}$ there are only two radiation zeros (the square root vanishes) and there are none for $E< E_{\mathrm crit.}$. Note that for $\beta=1$ the $W$s can be regarded as massless particles and, as in the case {\bf (a)} above, two zeros are present\footnote{$\tan \theta_{\gamma}= \frac{Q_q \sin \Theta}{1+Q_q \cos \Theta}$.}. But in practice $\beta<1$ and this gives rise to two additional zeros located close to the directions of the $W$s.  \\

\subsection{The general case}

In the previous section we have found radiation zeros in the soft-photon approximation. 
In order to extend these results to arbitrary
photon energies we have to consider the full matrix element, 
i.e. the sum of all the diagrams in Fig.~\ref{all}. Since we are interested now
in the case when $\Gamma_W \ll E_\gamma$, we can again make use of the partial 
fraction technique to factorise the full matrix element into
production and decay parts, exactly  as in Eqs.~(\ref{factorise}) and (\ref{sep}).
As in the previous section we focus on the $WW\gamma$ production process:
\bea
\label{crosssection}
d \, \sigma &= & \frac{1}{2s}  d\, \Phi_3 \,\, d \, \Phi_2^+ \,\, d \,{\Phi_2^-}
 \,\, \big  |\,\, {\cal M}_1 +  {\cal M}_2 +  
{\cal M}_3 +  {\cal M}_4 +  {\cal M}_9 \big. \no \\
&& \qquad \qquad +  {\cal M}_{10} +  {\cal M}_{11} +  {\cal M}_{12} + {\cal M}_{13} 
 + {\cal M}_{18}\,\, \big  | ^2 
\ea
where the subscript refers to the diagrams of Fig.~\ref{all}.\footnote{Note that
 only the first part of the partial fraction 
Eq.~(\ref{partial}) is to be taken for $ {\cal M}_3 , {\cal M}_4 , {\cal M}_{12}$ and $ {\cal M}_{13}$.} The final-state fermion parts of these diagrams are integrated over the  two-body phase
spaces to give two branching ratio ($W \to f \bar f$)  factors. The photon can be emitted off either the 
two initial-state quarks, the two final-state $W$'s,  
the internal lines  ($W$'s as well as the $t-$channel quark)
 or from the four boson vertex. 

\indent 
We next have to specify the three-body phase space configuration. 
To simplify the kinematics we choose
to fix the direction of the $W^-$ by
$\Theta$, 
and the energy and the angle of the photon by $E_{\gamma}$ and $\theta_{\gamma}$ respectively. 
An overall azimuthal angle is disregarded and, more importantly, the incoming and outgoing
particles are required to lie in a {\it plane}, defined by 
$\Phi=\phi_{\gamma}=0^{\circ}$.\footnote{We show later
that there are no radiation zeros for {\it non-planar} configurations.} 
Given the initial quark 
momenta $p_1,\, p_2$, the $W^+$ four-momentum is then constrained by 
energy-momentum conservation:
\bea
\label{kin2}
p_1^{\mu} & = & E(1,0,0,1) \no \\
p_2^{\mu} & = & E(1,0,0,-1) \no \\
k^{\mu} & = & E_{\gamma} (1, \sin\, \theta_{\gamma}, 0, 
\cos\, \theta_{\gamma}) \no \\
k_-^{\mu} & = & (E_W, \sqrt{E_W^2-M_W^2}\, \sin \, \Theta, 0, 
\sqrt{E_W^2-M_W^2}\, \cos \, \Theta) \no \\
k_+^{\mu} & = & (2E-E_{\gamma}-E_W, -( E_{\gamma} \sin\, 
\theta_{\gamma}+ \sqrt{E_W^2-M_W^2}\, \sin \, \Theta),0,    \no \\ 
&& \qquad\qquad\qquad   -( E_{\gamma} 
\cos\, \theta_{\gamma}+\sqrt{E_W^2-M_W^2}\, \cos \, \Theta)) 
\ea
where $E_W$ is determined by the constraint $ k_+^2=M_W^2$ and is given by
\begin{samepage}
\bea
E_W &=& \left\{ -2EE_{\gamma}^2 - 4E^3 +6E^2E_{\gamma}+ \left[ E_{\gamma}^2 \cos 
^2 (\theta_{\gamma}-\Theta) \left( -8E^3E_{\gamma} + 4E^2E_{\gamma}^2  \right.\right.\right. \no \\
&&\left. \left.\left. \hspace{0.5cm} +M_W^2E_{\gamma}^2 \cos ^2(\theta_{\gamma}-\Theta)
 + 4E^4-4E^2M_W^2+4EE_{\gamma}M_W^2-E_{\gamma}^2M_W^2 \right) \right] ^{\frac{1}{2}} \right\}\no \\
&& \hspace{1cm} /\left[E_{\gamma}^2 \cos ^2 (\theta_{\gamma} -\Theta) -4E^2 
+4EE_{\gamma}-E_{\gamma}^2\right] \quad .
\ea
\end{samepage}
In terms of these variables the three-body phase space integration is
\be
d\, \Phi_3(k_+,k_-,k)\, =\,   \frac{E_W E_\gamma}{4(2\pi)^5}\, 
 \frac{d\cos\Theta\, d\Phi \, \,
 dE_{\gamma} \, d\cos\theta_{\gamma}\, 
d\phi_{\gamma} \,
}{\left\vert -4E+2E_{\gamma}-2E_WE_{\gamma}\cos(\theta_{\gamma}-\Theta)
/\sqrt{E_W^2-M_W^2}\right\vert^2} \; .
\e

We first consider the differential cross section as a 
function of $\theta_{\gamma}$, with all  other variables kept fixed. 
For input parameters we take \cite{pdg}
\bea
\label{para1}
\begin{tabular}{lll}
$M_W =  80.41$ GeV, & $M_Z = 91.187$  GeV ,&  $ e^2={4 \pi}/{137.035}$, \\
$ g= {e}/{\sin \theta_w}$ ,& $ \sin^2 \theta_w = 0.23$ , & 
\end{tabular}
\ea
and in the following plots we also fix, for sake of illustration,
\be
\label{para2}
 Q_q=\frac{2}{3} \; , \qquad  E=500\;{\rm GeV} \; , \qquad  \Theta = \frac{2\pi}{3}  \; .  
\e

\begin{figure}[H]
\vspace{-2cm}
\centerline{\epsfysize=17cm\epsffile{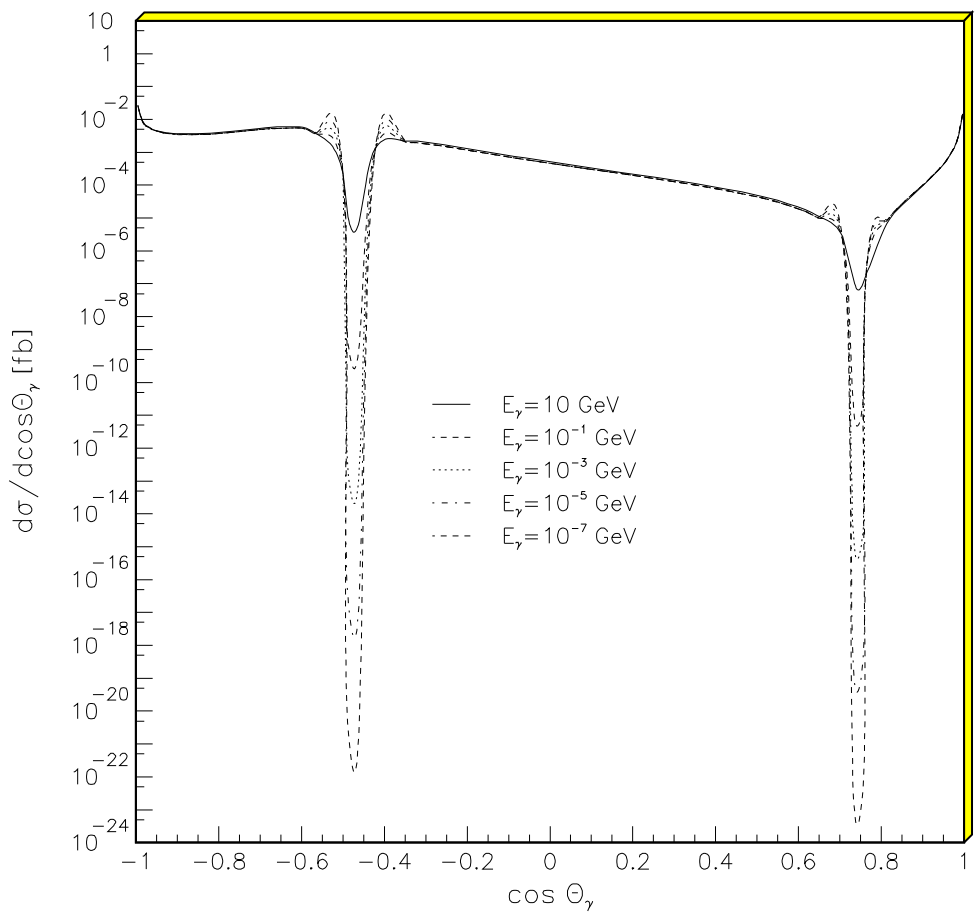}}
\vskip -2.5cm
\caption{\label{sm}{Differential cross section for the process
 $u \bar{u} \to W^+W^-\gamma $. }}
\end{figure}

Fig.~\ref{sm} shows the $\theta_{\gamma}$ dependence of the differential
$u \bar u \to W^+W^-\gamma$ cross section, for a selection 
of photon energies $E_{\gamma}$\footnote{Strictly, in order to separate out the $WW\gamma$ process in the first place we have to assume $E_{\gamma} \gg \Gamma_W$. 
However, to investigate the disappearance of the zero it 
is convenient to formally consider all $E_{\gamma}$ values down to zero.}. It is immediately
apparent that an actual zero of the cross section is only achieved in the limit $E_{\gamma}\to 0$.
Increasing the photon energy gradually `fills in' the dip. The reason is that for non-soft photons additional diagrams (9 and 18 in Fig.~\ref{all}) contribute and these give rise to a non-zero cross section at the positions of the zeros.\footnote{This is in contrast to the process $eq \to eq \gamma$ studied in Ref.~\cite{james} where {\it all} diagrams contribute in the soft-photon limit, and where the radiation zeros persist for $E_{\gamma} \neq 0$.}
The points at the bottom of the dips in Fig.~\ref{sm} are actually
the minimum values of the corresponding cross sections\footnote{The angles at which the minima occur are very close to the RAZ angle in the soft-photon limit, as can be seen in Fig.~\ref{sm}.}. In fact it can be shown that for
$E_\gamma$ not too large, $\sigma_{\rm min} \propto E_{\gamma}$. At high photon energies
the dips disappear altogether and the cross section assumes a different shape. \\

Note that the `zeros'/ dips both lie in the angular region between the outgoing $W^-$ and the incoming $u$, and by symmetry between the outgoing $W^+$ and the incoming $\bar{u}$. Further note that for a given $E_{\gamma}$, $\sigma_{\mathrm min}$ differs by a factor $\sim 100$ due to the asymmetric (with respect to the photon emission angle) contribution from diagram 18.\\

\begin{figure}[H]
\vspace{-1.7cm}
\centerline{\epsfysize=16.7cm\epsffile{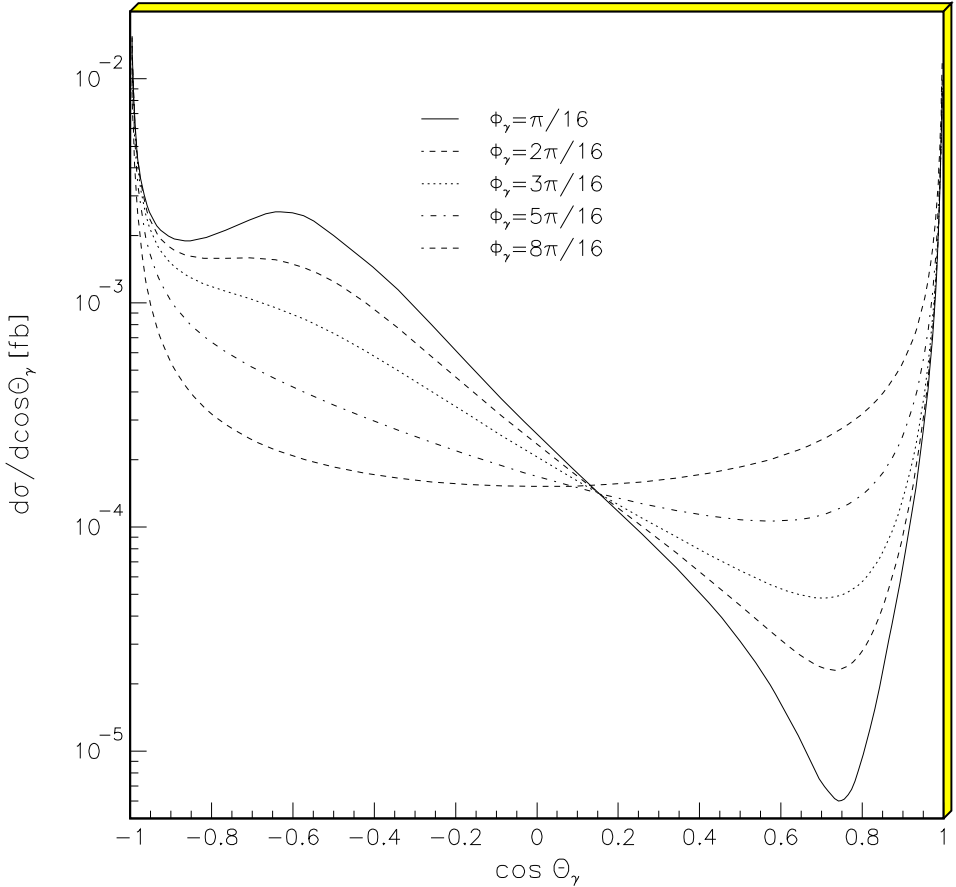}}
\vskip -2.2cm
\caption{\label{LRphi}{Same as Fig.~\ref{sm} for $E_{\gamma}=10^{-5}$~GeV and various $\phi_{\gamma}$.}}
\end{figure}

To confirm that we do indeed have a Type~II (planar configuration) radiation zero,
we next recalculate the $ \cos \theta_\gamma$ distribution for $\phi_\gamma \neq 0^\circ$. 
We choose a small non-zero photon energy $E_{\gamma}=10^{-5}$~GeV such that the dip is clearly
visible for $\phi_\gamma = 0^\circ$. The results are shown
in Fig.~\ref{LRphi}\footnote{Note the different scale compared to Fig.~\ref{sm}.}. For $\phi_\gamma$ well away from zero, there is no hint of a dip
in the cross section.

The results displayed in the above figures correspond to  $u\, \bar{u}$ scattering.
Similar results are obtained for  $d \, \bar{d}$ and $e^+ \, e^-$ scattering, i.e. 
exact Type~II zeros are only found in the soft-photon limit where they 
are given by Eq.~(\ref{tan}). The position of the zeros depends on the incoming
fermions' electric charge, and on the scattering angles and velocities of the $W$ bosons.
 For non-soft photons the dips are filled in, but still remain
clearly visible for photon energies up to ${\cal O}(10\; {\rm GeV})$.

\section{Anomalous gauge boson couplings}

As discussed in the Introduction, the existence of radiation zeros is in general
destroyed by the presence of anomalous gauge boson couplings. Two categories of such
couplings are usually considered --- 
trilinear and quartic gauge boson couplings  --- and each probes different 
aspects of the weak interactions. The trilinear couplings directly  test
the
non-Abelian gauge structure, and possible deviations from the SM
forms have been extensively studied
in the literature, see for example \cite{tgvtheory} and references therein. 
Experimental bounds have also been obtained \cite{tgvexpt}. 
In contrast, the quartic couplings 
can be regarded as a more direct window on electroweak symmetry breaking,
in particular to the scalar sector of the theory (see for example
\cite{godfrey}) or, 
more generally, on new physics which couples to electroweak bosons. 
In this respect it is quite possible that the quartic couplings deviate
from their SM values 
while the triple gauge vertices do not. For example,
if the mechanism for electroweak symmetry breaking does not reveal itself
through 
the discovery of new particles such as the Higgs 
boson, supersymmetric particles or technipions,  it is  
possible that anomalous quartic couplings could provide the first evidence 
of new physics in  this sector of the electroweak theory \cite{godfrey}.
\\

The impact of anomalous trilinear couplings on the radiation zeros in the 
$q \bar q' \to W \gamma$ process was considered in Ref.~\cite{baur}.
 As expected, the zeros
are removed for non-zero values of the anomalous parameters. Such couplings would also affect the zeros in the $WW\gamma$ case. However there are already quite stringent limits on these trilinear couplings from the Tevatron $p\bar{p} \to W\gamma X$ \cite{tgctev} and LEP2 $e^+e^- \to W^+W^-$ processes \cite{tgvexpt}. We therefore neglect them here and concentrate on genuine anomalous quartic couplings, for which no limits exist at present. The $WW\gamma$ process is in fact
the simplest one which is sensitive to {\it quartic} couplings. It is natural therefore
to consider the implications of anomalous quartic couplings on the radiation zeros
discussed in the previous section. \\

The lowest dimension operators which lead to genuine quartic couplings 
where at least one photon is involved are of dimension 6 \cite{belanger}. First,  we
have  the neutral
and  charged Lagrangians, both giving anomalous contributions
to the $VV\gamma\gamma$ vertex, with $VV$ either being $W^+W^-$ or
$Z^0Z^0$.
\bea
\label{L0}
{\cal L}_0 &=& - \frac{e^2}{16 \Lambda^2}\, a_0\, F^{\mu \nu} \, F_{\mu
\nu} \vector{W^{\alpha}} \cdot \vector{W_{\alpha}} \no \\
&=&  - \frac{e^2}{16 \Lambda^2}\, a_0\, \big[ - 2 (k_1 \cdot k_2 ) 
( A \cdot A) + 2 (k_1 \cdot A)(k_2 \cdot A)\big] \no \\
&& \hspace{1.5cm} { \times} \big[ 2 ( W^+ \cdot W^-) +  (Z \cdot Z) /
\cos ^2 \theta_w \big]  \quad ,
\ea
\bea
\label{Lc}
{\cal L}_c &=& - \frac{e^2}{16 \Lambda^2}\, a_c\, F^{\mu \alpha} \, F_{\mu
\beta} \vector{W^{\beta}} \cdot \vector{W_{\alpha}} \no \\
&=& - \frac{e^2}{16 \Lambda^2}\, a_c\, \big[- (k_1 \cdot k_2)\, A^{\alpha}
A_{\beta} +(k_1 \cdot A)\, A^{\alpha} k_{2 \beta} \big. \no \\
&& \hspace{1.8cm}\big. \quad \quad + (k_2 \cdot A)\,
k_1^{\alpha} A_{\beta} -(A \cdot A)\, k_1^{\alpha} k_{2 \beta} \big] \no
\\
&& \hspace*{1.5cm} { \times} \big[ W_{\alpha}^- W^{+ \beta} +
W_{\alpha}^+ W^{-
\beta} + Z_{\alpha} Z^{\beta} / {\cos ^2 \theta_w} \big] \ . 
\ea
where $k_1$ and $k_2$ are the photon momenta.
Since we are interested in the anomalous $WW\gamma\gamma$ contribution 
we select the corresponding part of the Lagrangian. \\ 

Second, an anomalous $WWZ\gamma$ vertex is obtained from the Lagrangian
\bea
\label{Ln}
{\cal L}_n &=&  - \frac{e^2}{16 \Lambda^2}\, a_n \epsilon_{ijk} W_{\mu
\alpha}^{(i)} W_{\nu}^{(j)} W^{(k)\alpha} F^{\mu \nu} \no \\
&=&  - \frac{e^2}{16 \Lambda^2 \cos \theta_w} \, a_n \, \big( k^{ \nu}
A^{\mu} - 
k^{\mu} A^{\nu} \big) \no \\
&& { \times}  \left( - W^-_{\nu} k_{\mu}^+\,  ( Z \cdot W^+) +
W_{\nu}^+
k_{\mu}^-\, ( Z \cdot W^-) + Z_{\nu} k_{\mu}^+\, (W^+ \cdot W^-) \right.
\no \\
&& - Z_{\nu} k_{\mu}^-\, (W^+ \cdot W^-) + W_{\nu}^- W_{\mu}^+\, (k^+
\cdot Z) - W_{\nu}^+ W_{\mu}^- \,(k^-  \cdot Z) \no \\
&& - Z_{\nu} W_{\mu}^+\, (k^+ \cdot W^-) + Z_{\nu}W_{\mu}^-\, ( k^- \cdot
W^+) - W^+_{\nu} k^0_{\mu} \,(Z \cdot W^-) \no \\
&& \left. + W_{\nu}^- k^0_{\mu} \, (Z \cdot W^+) - W_{\nu}^- Z_{\mu}\,
(k^0 \cdot W^+) +W_{\nu}^+ Z_{\mu}\, (k^0 \cdot W^-) \right) 
\ea
where 
 $k, k^+, k^-$ and $k^0$ are the momenta of the photon, $W^+$, $W^-$
and $Z^0$ respectively.\\

Let us consider first the differential cross section in the planar configuration
as  a function of $\theta_{\gamma}$, just as we did in the previous
section, but now in the presence of non-zero values of the three anomalous
parameters $a_0,\, a_c$ and $a_n$ introduced above. 
From the Lagrangian it can be seen that any
anomalous contribution is {\it linear} in $E_{\gamma}$. Soft photons are `blind'
to the anomalous couplings and therefore the zeros in the $k^{\mu} \to 0$ limit
survive. For moderate photon energies, the dips in the SM cross section
will be filled in by contributions proportional to $a_i$ and $a_i^{\phantom{i} 2}$. 
The higher the photon energy, the more dramatic the effect, although of course
the dips become less well defined too.
In principle, therefore, one should optimize the photon energy, to make it small enough
to maintain the zeros but at the same time
large enough to gain measurable deviations from the SM prediction. Since the anomalous
contributions originate in the four boson vertex in the $s$-channel,
 one can also increase the sensitivity to them by
considering only right-handed initial quarks, for which  the $t$-channel contributions
are absent.\footnote{Unfortunately in doing so
one also decreases the total cross section by roughly 2 orders of magnitude, so 
again it has to be seen whether the sensitivity to new physics is in fact increased
in practice.}\\

\begin{figure}[H]
\vspace{-2.7cm}
\centerline{\epsfysize=17cm\epsffile{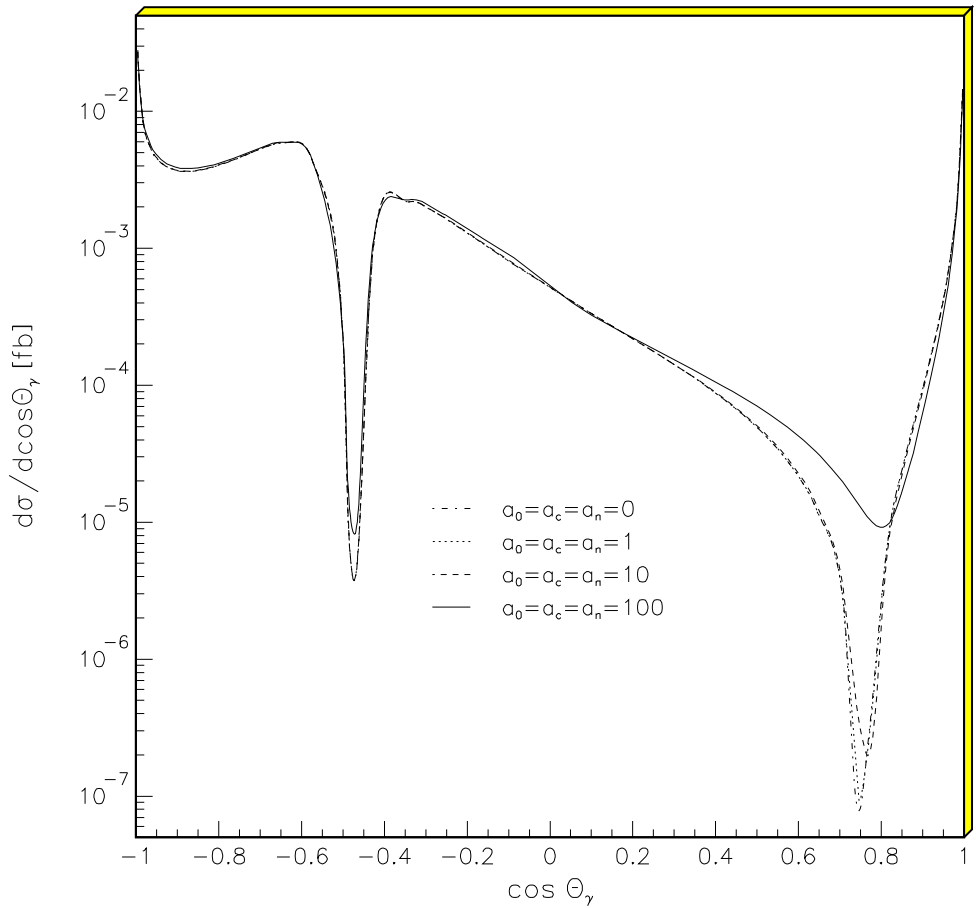}}
\vspace{-2.5cm}
\caption{\label{bsm}{Differential cross section for the 
process $ u \bar{u} \to W^+W^-\gamma 
$ 
with $E_{\gamma}=1$~GeV. The curves correspond to different values
of the anomalous parameters introduced in the text.}}
\end{figure} 

Fig.~\ref{bsm} shows the $\theta_\gamma$ dependence of the $u \bar u \to WW\gamma$
cross section for $E_\gamma = 1$~GeV, for the same configuration and parameters
as in Fig.~\ref{sm} (see Eqs.~(\ref{para1},\ref{para2})). 
The curves correspond to different (positive) values of the anomalous 
parameters
\footnote{To make quantitative predictions the anomalous 
parameter $\Lambda$ appearing in Eqs.~(\ref{L0},\ref{Lc},\ref{Ln}) has to be fixed. 
We choose $\Lambda=M_W$; any other choice  results in a trivial rescaling of the 
anomalous parameters $a_0, a_c$ and $a_n$. See the discussion in 
\cite{sw1}. The anomalous parameters can also be negative, which leads to
results similar to those in  Fig.~\ref{bsm}.}. The anomalous contribution is approximately isotropic in $\theta_{\gamma}$. Therefore because the dips have different depths one gets filled in more rapidly than the other. This is evident in the figure, where the dip at $\cos \theta _{\gamma} \sim 0.75$ is already filled in for $a_i \sim {\cal O}(100)$, whereas the steep dip at $\cos \theta_{\gamma} \sim -0.48$ is still very apparent. This shows it is advantageous to focus on certain regions of photon phase space in order to increase the sensitivity to the anomalous couplings. Of course, this requires very high luminosity to ensure a large enough event rate in these regions.\\

\section{Conclusions} 

We have investigated the (Type~II) radiation zeros of the process 
$ q \bar{q} \to W^+W^-\gamma \to f_1 \bar{f}_2 f_3 \bar{f}_4 \gamma $. 
In the soft-photon limit ($\Gamma_W \ll E_{\gamma} \ll 1$) the cross section vanishes
for certain values of the photon and $W^\pm$ production angles, for which analytic
expressions have been derived (Eq.~(\ref{tan})).
For non-zero photon energies the zeros disappear, but for energies not too large
the photon angular distribution still exhibits deep dips centred on the positions of the soft-photon zeros. 
The subtle cancellations leading to the zeros in the soft-photon limit still takes place, but for non-soft photons two additional diagrams (9 and 18) have to be considered and $\sigma_{\mathrm min}$ is exactly that contribution. In the `classic` process $q \bar {q}\,' \to W^+ \gamma$ there are no further diagrams for non-soft photons and the zeros survive for all photon energies.
Although we have concentrated on the quark scattering process $q \bar {q} \to W^+W^- \gamma$ our results apply equally well to $e^+e^- \to W^+W^-\gamma$ by setting $Q_q=-1$. Note that, as for quark antiquark scattering, the zeros in the $e^+e^-$ (soft-photon) case are in the `visible' regions of phase space, in contrast to those in the analogous `classical' process $ e^+\nu_e \to W^+ \gamma$. Furthermore for $e^+e^-$ scattering diagram 9 (scattering via $t$-channel exchange) is not present and direct access to the four boson vertex is given for non-soft photons, i.e. the four boson vertex is the only contribution to the cross section at the position of the zeros.\\

We have also studied the effect of including non-zero anomalous quartic couplings.
These contributions increase with increasing photon energy and fill in the dips present in the standard model.
In principle, therefore,
the vicinity of the radiation zeros is the most 
sensitive part of phase space to these anomalous four boson couplings. \\

Our analysis has been entirely theoretical. Having established that there {\it are}
regions of phase space where  the cross section is heavily suppressed, the next step
is to see to what extent the phenomenon persists when hadronisation, radiative corrections, smearing, boost, detector
etc. effects are taken into account, in the context, for example,
of a possible measurement at the Tevatron or LHC  hadron colliders. In this respect, a high energy, high luminosity $e^+e^-$ linear collider could provide a cleaner environment for studying $WW\gamma$ production in this way.

\newpage
\vspace*{-0.6cm}
\noindent{\bf Acknowledgements}\\ 
This work was supported in part by the EU Fourth Framework Programme 
`Training and Mobility of 
Researchers', Network `Quantum Chromodynamics and the Deep Structure of 
Elementary Particles', 
contract FMRX-CT98-0194 (DG 12 - MIHT). AW gratefully acknowledges financial 
support in the form of a `DAAD Doktorandenstipendium im Rahmen des gemeinsamen 
Hochschulprogramms III f\"ur Bund und L\"ander'.\\


\vspace*{-0.7cm}
\begin{thebibliography}{99}
\bibitem{first} K.O. Mikaelian, M.A. Samuel and D. Sahdev, \prl{43}{79}{746}.
\vspace{-0.17cm}
\bibitem{brown} R.W. Brown, {\it Understanding Something about Nothing: Radiation Zeros},
published in Vector Boson Symp.~1995:~261-272. \\
R.W. Brown, K.L. Kowalski and S.J. Brodsky \prd{28}{83}{624}.
\vspace{-0.17cm}
\bibitem{cdf} D. Benjamin for the CDF collaboration, 
{\it $W\gamma $ and $Z \gamma$ production at the Tevatron}, FERMILAB-Conf-95-241-E.
\vspace{-0.17cm}
\bibitem{kleiss} R. Kleiss and W.J. Stirling, \np{262}{85}{235}.
\vspace{-0.17cm}
\bibitem{tt} V.A. Khoze, W.J. Stirling and L.H. Orr, \np{378}{92}{413}.
\vspace{-0.17cm}
\bibitem{orr} Yu.L. Dokshitzer, V.A. Khoze, L.H. Orr and W.J. Stirling, \pl{313}{93}{171}.
\vspace{-0.17cm}
\bibitem{james} M. Heyssler and W.J. Stirling, \epj{4}{98}{289}.
\vspace{-0.17cm}
\bibitem{pdg} C. Caso {\it et al.}, {\it Review of Particle Physics}, \epj{3}{98}{1}.
\vspace{-0.17cm}
\bibitem{tgvtheory} K.~Hagiwara, R.D.~Peccei, D.~Zeppenfeld and K.~Hikasa,
\np{282}{87}{253}. \\
{\it Triple Gauge Boson Couplings},
G.~Gounaris {\it et al.}, 
in `Physics at LEP2', Vol.~1, p.~525-576, CERN (1995) [hep-ph/9601233]. 
\vspace{-0.17cm}
\bibitem{tgvexpt} ALEPH Collaboration:  R.~Barate {\it et al.}, \pl{422}{98}{369};
preprint CERN-EP-98-178, November 1998 [hep-ex/9901030]. \\
OPAL Collaboration: G.~Abbiendi {\it et al.}, preprint CERN-EP-98-167,
October 1998 [hep-ex/9811028].
\vspace{-0.17cm}
\bibitem{godfrey} S. Godfrey, {\it Quartic Gauge Boson Couplings}, 
published in  Proceedings of the International Symposium on Vector Boson Self-Interactions, 
UCLA, Feb.~1-3, 1995.
\vspace{-0.65cm}
\bibitem{baur} T. Abraha and  M.A. Samuel,  Oklahoma State U.  preprint OSU--RN--326,
hep-ph/9706336.
\vspace{-0.17cm}
\bibitem{tgctev}H.T. Diehl, {\it Boson Pair Production and Triple Gauge Couplings}, FERMILAB-CONF-97-216-E. 
\vspace{-0.17cm}
\bibitem{belanger} G.B$\acute{{\rm e}}$langer and F. Boudjema, \pl{288}{92}{201}.
\vspace{-0.17cm}
\bibitem{sw1} W.J. Stirling  and A. Werthenbach, preprint hep-ph/9903315.

\end{thebibliography}
\end{document}